\documentclass[preprint,aps,prc,showpacs,nofootinbib]{revtex4}
\usepackage{epsfig}
\usepackage{graphicx,amsmath}

\newcommand\ba{\begin{eqnarray}}
\newcommand\ea{\end{eqnarray}}
\newcommand\be{\begin{equation}}
\newcommand\ee{\end{equation}}
\newcommand\nn{\nonumber}

\def\phi{\varphi}

\def\be{\begin{equation}}
\def\ee{\end{equation}}
\def\ba{\begin{eqnarray}}
\def\ea{\end{eqnarray}}

\begin{document}
Electron-positron pair production by linearly
polarized photon in the nuclear field.

  S. Bakmaev$^a$, E.~A.~Kuraev$^a$, I. Shapoval$^b$ and \fbox{Yu.~P.~Peresun'ko$^b$}
.\\
a)JINR,\quad Dubna,\quad Russia; b)KIPT,\quad Kharkov,\quad Ukraina.

\section{Abstract}
Process of lepton pair production by polarized photon can be used
to measure the degree of linear polarization of high energy photon.
The differential cross section and the analyzing power are calculated
with taking into account higher powers of expansion on $Z\alpha$.
Pure Coulomb and screened potential are considered.

\section{Introduction}
Studying of process of pair creation starts from celebrated papers of 1953-1969
\cite{Hull,Bethe,Olsen,Maximon,Motz,Olsen2}
and continue to attract the attention up to now. Main interests
nowadays is to use this process as a polarimeter. Really it have rather large
cross section and the polarization effects can reach $14$ percents,\cite{Mork}. Two different
mechanisms
of pair creation must be taken into account: the Bethe-Heitler one, when the pair
is produced in collision of two photon-one real and other virtual and the
bremsstrahlung mechanism,when pair is created by single virtual photon. It was shown
in fundamental papers of E.Haug \cite{Haug} that at photon energies exceeding $50MeV$ in the
target Laboratory frame the contribution of bremsstrahlung mechanism as well as
the interference of corresponding amplitude with two-photon ones do not exceed 5
percents and decreased with further photon energy growth. Taking into account the
lowest order radiative corrections (RC) do not change the situation. For the case
of target such as
proton or light nuclei the main contribution to (RC) are connected with final state
interaction between pair components. Two virtual photon exchange between pair and
nuclei amplitude do not interfered with Born amplitude as well as they have different
signatures. Pure two photon exchange amplitude contribution do not contain any
enhancement factors such as "large logarithms" of ratio of photon energy $\omega$
to lepton mass $m$,and have of order $\alpha^2$. It can be neglected compared with
contribution of order $\alpha/\pi$ coming from interference of Born amplitude with
1-loop ones connected with lepton pair interaction.

The situation changed when one consider the
pair creation on heavy nuclei with charge parameter $\nu=Z\alpha$ not too small.
Main contribution arises from many photon exchanges mechanism between pair component
with nuclei.

Total cross section of pair creation process
by photon on a nuclei of charge $Z$
\ba
\gamma(k)+Y(P,Z) \to e^+(p_+)+e^-(p_-)+Y(P',Z),\quad q=P-P',\quad s=2Pk=2M\omega,\\ \nn
P^2=(P')^2=M^2,\quad p_\pm^2=m^2,\quad |q^2|\ll s,
\ea
for the case of unpolarized photon is \cite{Akhiezer}
\ba
\sigma=\frac{28}{9}\frac{Z^2\alpha^3}{m^2}[\ln\frac{2\omega}{m}-\frac{109}{42}-f(\nu)], \\ \nn
f(\nu)=\nu^2\sum_{n=1}^\infty\frac{1}{n(n^2+\nu^2)}.
\ea
This result was recently reproduced in informative paper of Ivanov an Melnikov \cite{Ivanov},
where the differential cross section was as well considered.

Direction of $e^+$ and $e^-$ emitting correlates with degree and direction of photon linear
polarization and so this process can serve as polarimetric reaction for the measuring
linear polarization of high energy photons. At the present beams in the region of photons.
At the present time experimental tasks requests measurements of linear polarization of photon beams
in the region of photon energy up to 1-2 Gev with accuracy near 1 percent or better. So need
 to know cross section of $e^+e^-$ pair photo-production and analyzing power of this
appropriated accuracy.

The considered process was discussed in details in the works mentioned above.
Calculations in these works was carried out on the base of wave functions of the final
electron and positron in the external screened Coulomb field in Furry-Sommerfeld-Maue
approximation. This approximation is valid for high energy of produced particles,
$m/\epsilon_{\pm}\ll 1$, and for small emitting angles $\theta_{\pm}\sim m/\epsilon_{\pm}$
 where $\epsilon_{\pm}\leq10^{-3}$, is energy of positron or electron, $m$ is electron mass.

It is well known that main contribution to cross section of considered process gives just
the region of small emitting angles. But it should be noted that for that for the purposes
of high-energy photon polarimetry, where restriction on the value of observable in the
experiment angles is $\theta^*_{\pm} $
In this paper we use formalism of \cite{Ivanov} paper to consider the case of linearly polarized
photon.

First we briefly sketch the relevant results of paper \cite{Ivanov}.
Sudakov's parameterization of 4-momenta is used below:
\ba
q=\alpha_q k+\beta_q \tilde{P}+\vec{q}, \\ \nn
q_i=\alpha_i k+\beta_i \tilde{P}+\vec{q}_i, \\ \nn
p_+=x_1 k+y_1 \tilde{P}+\vec{p}_1, \\ \nn
p_-=x_2 k+y_2 \tilde{P}+\vec{p}_2,
\ea
with $\vec{a}$-euclidean two dimensional vector $\vec{a}=(0,0,a_x,a_y)$, orthogonal to photon 4-momentum
$k=\omega(1,1,0,0)$,
$\tilde{P}=(M/2)(1,-1,0,0)=P-k(M^2/s)$ is light-like 4-vector. The conservation low and
on mass shell conditions leads to
\ba
x_1+x_2=1,\quad y_1=\frac{c_1}{x_1s},\quad y_2=\frac{c_2}{x_2s}, \nn \\
c_l=\vec{p}^2_l+m^2, l=1,2; \quad \vec{q}=\vec{p_1}+\vec{p_2}. \ea
Matrix element corresponding to $N$ photon exchange is \ba
M_N=-i^Ns\frac{8\pi^2(eZ)^N}{N!}\int\Pi_{i=1}^N\frac{d^2q_i}{(2\pi)^2}\frac{F(q_i^2)}{\vec{q}_i^2}
\delta^2(\sum q_i-q)J^N_{\gamma\to l\bar{l}}, \ea where
$J^N_{\gamma\vec{l}\bar{l}}$- are impact factor which rewritten in
the simple form \cite{Beres,Ivanov,Kuraev} with \ba J^N_{\gamma\to
l\bar{l}}(\vec{p}_1,\vec{p}_2)=\bar{u}(p_-)[mS^N\hat{\epsilon}-
2x_1\vec{T}^N\vec{\epsilon}-\hat{T}^N\hat{\epsilon}]
\frac{\hat{\tilde{P}}}{s}v(p_+). \ea The quantities $S,\vec{T}$ obey
the recurrent relations \ba
S^N(\vec{p}_1,\vec{p}_2,\vec{q}_N)=S^{N-1}(\vec{p}_1,\vec{p}_2-\vec{q}_N)-
S^{N-1}(\vec{p}_1-\vec{q}_N,\vec{p}_2),N=2,3,... , \ea and the
similar expression for $\vec{T}^N$. The initial values, which
corresponds to one photon exchange are \ba
S^1=S^1(\vec{p}_1,\vec{p}_2)=\frac{1}{c_1}-\frac{1}{c_2}; \quad
\vec{T}^1=\vec{T}^1(\vec{p}_1,\vec{p}_2)=\frac{\vec{p}_1}{c_1}+
\frac{\vec{p}_2}{c_2}. \ea Introducing the values \ba
J^{(N)}_{S,\vec{T}}=\int
\Pi_1^N\frac{d^2q_iF(q_i^2)}{\vec{q}_i^2}[S^N,\vec{T}^N]\delta^2(\sum
q_i-q), \ea and their Fourier transform: \ba
I^{(N)}_{S,\vec{T}}(\vec{r}_1,\vec{r}_2)=\int\frac{d^2\vec{p}_1d^2\vec{p}_2}{(2\pi)^2}
e^{i\vec{p}_1\vec{r}_1+i\vec{p}_2\vec{r}_2}J^{(N)}_{S,\vec{T}}, \ea
the recurrent relations can be written in form: \ba
I^{(N)}_{S,\vec{T}}(\vec{r}_1,\vec{r}_2)=\pi
I^{(N-1)}_{S,\vec{T}}(\vec{r}_1,\vec{r}_2)
\Phi(\vec{r}_1,\vec{r}_2), \\ \nn
\Phi(\vec{r}_1,\vec{r}_2)=\frac{1}{\pi}\int(e^{i\vec{q}\vec{r}_2}-e^{i\vec{q}\vec{r}_1})
\frac{d^2qF(q^2)}{\vec{q}^2}. \ea In Moliere approximation of atomic
form-factor in Tomas-Fermi model (we use it below) the expression
for form-factor is \cite{Moliere}: \ba
\frac{F(q^2)}{\vec{q}^2}=\frac{1-F_A}{\vec{q}^2}=\sum_1^3\frac{\alpha_i}{\mu_i^2+\vec{q}^2},
\ea with $\alpha_1=0.1;\alpha_2=0.55;\alpha_3=0.35$ and
$\mu_i=(mZ^{1/3})b_i$ with $b_1=6.0;b_2=1.2;b_3=0.3$. For this case
the analytic expressions con be obtained: \ba
\phi_{12}=\Phi(\vec{r}_1,\vec{r}_2)=2\sum_1^3\alpha_i[K_0(\mu_i|r_2|)-K_0(\mu_i|r_1|)].
\ea For pure Coulomb potential $F(q^2)=1$, we have
$\Phi^c(\vec{r}_1,\vec{r}_2)=\ln\frac{\vec{r}_1^2}{\vec{r}_2^2}$.

Boundary of recurrent relations are
\ba
I_S^{(1)}(\vec{r}_1,\vec{r}_2)=\frac{1}{2}K_0(m|\vec{r}_1-\vec{r}_2|)\Phi(\vec{r}_1,\vec{r}_2); \\ \nn
\vec{I}_T^{(1)}(\vec{r}_1,\vec{r}_2)=\frac{Im(\vec{r}_1-\vec{r}_2)}{2|\vec{r}_1-\vec{r}_2|}
K_1(m|\vec{r}_1-\vec{r}_2|)\Phi(\vec{r}_1,\vec{r}_2),
\ea
with $K_{0,1}(z)$-modified Bessel functions. The summation on the number of exchanged photons can be performed:
\ba
J_S(\vec{p}_1,\vec{p}_2)=\frac{i}{2\nu}\int\frac{d^2r_1d^2r_2}{(2\pi)^2}e^{-i\vec{p}_1\vec{r}_1-i\vec{p}_2\vec{r}_2}
K_0(m|\vec{r}_1-\vec{r}_2|)[e^{-i\nu\Phi(\vec{r}_1,\vec{r}_2)}-1]; \\ \nn
J_T(\vec{p}_1,\vec{p}_2)=\frac{-1}{2\nu}\int\frac{d^2r_1d^2r_2}{(2\pi)^2}e^{-i\vec{p}_1\vec{r}_1
-i\vec{p}_2\vec{r}_2}\frac{Im(\vec{r}_1-\vec{r}_2)}{2|\vec{r}_1-\vec{r}_2|}
K_1(m|\vec{r}_1-\vec{r}_2|)[e^{-i\nu\Phi(\vec{r}_1,\vec{r}_2)}-1].
\ea
Differential cross section have a form:
\ba
d\sigma=\frac{2\alpha\nu^2}{\pi^2}[|\vec{J}_T|^2+m^2|J_S|^2-4x(1-x)\vec{J}_T\vec{\epsilon}
\vec{J}_T\vec{\epsilon^*}]dxd^2p_1d^2p_2=\\ \nn
\frac{2\alpha\nu^2}{\pi^2}[W_{unp}+\xi_3W_{pol}\cos(2\phi)]dxd^2p_1d^2p_2,\\
e_ie^*_j\to\frac{1}{2}[I+\xi_1\sigma_1+\xi_3\sigma_3]_{ij}; \quad i,j=x,y.
\ea
with polarization degree of photon, described by means of $\xi_{1,3}$-it's Stokes parameters,
$\phi$ is the angle between the vector $\vec{J}_T$ and the direction of maximal polarization of
photon (if we choice the $x$ axes along the direction of maximal polarization of photon,
we put $\xi_1=0;P=\xi_3$), and
\ba
W_{unp}=[x^2+(1-x)^2]|\vec{J}_T|^2+m^2|J_S|^2; \\ \nn
W_{pol}=-2x(1-x)|\vec{J}_T|^2.
\ea
For the case of screened potential (ignoring the experimental conditions of pairs component
detection) we use the expression for phase given above.
Performing the integration on pair momenta we obtain:
\ba
2\pi\frac{d\sigma}{dxd\phi_1}=\frac{2\alpha}{m^2}\int\limits^{2\pi}_0\frac{d\phi}{2\pi}\int\limits_0^\infty
dx_1\int\limits_0^\infty dx_2(1-\cos(\nu\phi_{12}))[K_0^2(z)+[x^2+(1-x)^2]K_1^2(z)- \\ \nn
2x(1-x)\xi_3K_1^2(z)\cos(2\phi_1)], z=\sqrt{x_1+x_2-2\sqrt{x_1x_2}\cos\phi},
\ea
with the azimuthal angle $\phi_1$ is the angle between the direction of maximal photon polarization
and the plane containing the direction of initial photon and electron (positron) from the
pair.

For the case of pure Coulomb potential integration in (19) diverge
and must be regularized. We leave here this academic problem. For
the case of screened potential we obtain: \ba \label{20}
2\pi\frac{d\sigma}{dxd\phi_1}=\frac{2\alpha}{m^2}[a(\nu)+(x^2+(1-x)^2)b(\nu)-2x(1-x)\xi_3\cos(2\phi_1)b(\nu)].
\ea The $\nu$ dependence of coefficients $a(\nu),b(\nu)$ is shown
in Fig. 1.

Further we consider the realistic case of nonzero momentum,
transferred to nuclei $|\vec{q}|^2\gg m^2_e=m^2$. For the case of
pure Coulomb potential we have \ba
\frac{d\sigma}{dxd\Omega_+d\Omega_-}=\frac{2\alpha\nu^2\omega^4}{\pi^2m^2}\frac{x^2(1-x)^2}{(\vec{q}^2)^2}
|\Gamma(1-i\nu)|^4[W_u^c+\xi_3W_p^c\cos(2\phi)], \ea with \ba
\label{Eq21} W_u^c=m^2[x^2+(1-x)^2]|(2F_1-F_2)\frac{\vec{p}_2}{c_2}
+F_2\frac{\vec{p}_1}{c_1}|^2+ \\ \nn
|F_2-F_1+(2F_1-F_2)\frac{m^2}{c_2} -F_2\frac{m^2}{c_1}|^2; \\ \nn
W_p^c=-2x(1-x)m^2[|2F_1-F_2|^2\frac{\vec{p}_2^2}{c_2^2}\cos(2\phi_+)+
\\ \nn +|F_2|^2\frac{\vec{p}_1^2}{c_1^2}\cos(2\phi_-)+
2Re(F_2^*(2F_1-F_2))\frac{|\vec{p}_2|}{c_2}\frac{|\vec{p}_1|}{c_1}\cos(\phi_++\phi_-)],
\ea with
$F_1=F(i\nu,-i\nu;1;z);F_2=(1-i\nu)F(i\nu,1-i\nu;2;z)$-hypergeometric
functions, $z=1-(m^2\vec{q}^2/[c_1 c_2])$, the value of transverse
component of pair are
$$|\vec{p}_1|=\omega x\theta_-;|\vec{p}_2|=\omega (1-x)\theta_+ ,$$
$x,1-x$ are the energy fractions of electron, positron. Azimuthal angles
$\phi_\pm$ are the angles between the direction of maximal polarization of photon and
the transverse component of positron and electron; $d\Omega_+d\Omega_-=d\theta_+d\phi_+d\theta_-d\phi_-$
with $\theta_i,\phi_i$-are the polar and azimuthal angles of pair component emission.

For the case of small transferred to nuclei momentum
$m^2\ll\vec{q}^2\ll\vec{p}_1^2\approx\vec{p}_2^2$ we can put in
(\ref{Eq21})$z=1$ and using $F_1=F_2=|\Gamma(1-i\nu)|^{-2}$
\cite{GR}, we reproduce the cross section in Born approximation \ba
\label{Eq22} \frac{d\sigma^c_B}{dx
d\Omega_+d\Omega_-}=\frac{2\alpha^3
Z^2\omega^4x^2(1-x)^2}{\pi^2(\vec{q}^2)^2}
\times\left[m^2 (S^1)^2+(x^2+(1-x)^2)(\vec{T}^1)^2\right.\nn \\
\left.-2x(1-x)\xi_3 (\frac{\vec{p}_2^2}{c_2^2}\cos(2\phi_2)+
+\frac{\vec{p}_1^2}{c_1^2}\cos(2\phi_1)+2\frac{|\vec{p}_2|}{c_2}\frac{|\vec{p}_1|}{c_1}\cos(\phi_1
+\phi_2))\right]. \ea And the similar expression for the case of
screened potential with the replacement \ba \frac{1}{\vec{q}^2} \to
\sum_1^3\frac{\alpha_i}{\mu_i^2+\vec{q}^2}. \ea We note that the
quantity in square brackets in the rhs of (\ref{Eq22}) equation is
proportional to $\vec{q}^2$ at small $\vec{q}^2$. The experimental
restrictions connected with pair component detection can be imposed
as a domain of variation of energy fractions and angles of electron
and positron.

In Appendix we give the explicit expression of two virtual photons exchange to matrix element.
The limit of large transversal momenta is also considered.

\section{Discussion}

In a famous papers of Bethe, Maximon and Olsen [2-4] the general theory of pair production and
bremsstrahlung was build basing on on the knowledge of electron wave function in Coulomb field.
Part of these results were reproduced in perturbation theory in [9]. Unfortunately the expression
for the differential cross sections which can be used in current experiments with some cuts were
rather poorly presented. It is the motivation of this paper.
Part of distributions, needed for experiment we derive above. It is the differential distribution
on energy fractions and the emission angles of both pair component for the case of unscreened
potential (21),(22); the inclusive distribution on one of pair components for the case of screened
potential (19),(20). As well we had investigate the case of large transversal momenta of pair, and
show that all the Coulomb corrections disappears in this limit. These results are new and provide
high accuracy since they are valid in all orders of perturbation theory. Distribution (21) allows
to put experimental cuts in set-up with both electron and positron tagged. The dependence on
$Z\alpha$ parameter in distribution on energy fraction and the azimuthal angle is shown in Fig.1.

Asymmetry calculated by formulae (20) (see Fig.2) \ba
A_1(x,\xi_3=1)=\frac{d\sigma(\phi_1=0,x)-d\sigma(\phi_1=\pi/,x)}
{d\sigma(\phi_1=0,x)+d\sigma(\phi_1=\pi/2,x)}, \ea is as well large
compared with the results obtained in [3]. The reason: is in [3] an
restriction on emission angles was $\theta_\pm>10^{-3}$ put on. The
main contribution which is relevant in (20), arises from the
emission angles much more smaller corresponding to $p_{1,2}<<m$.

Results for cross sections given above do not depend on photon energy which in accordance with the results
obtained in paper [13]. Namely the other mechanisms of pair creation give negligible contribution (on the level
of several percents compared with the one considered above) starting from $\omega>50 MeV$.

To estimate the order of magnitude of cross section we calculate it for unpolarized case for $Z=79$ to be
$\sigma=170 mb$.

The accuracy of our calculations is determined by the omitted terms
\be
1+O(\frac{p^2}{s},\frac{\alpha}{\pi}\ln\frac{p^2}{m^2}).
\ee
The quantity of errors is of order of several percents. The last term corresponds to the final state interaction
of the pair component, which was not considered here.

\begin{figure}
\begin{center}
\includegraphics[width=12cm]{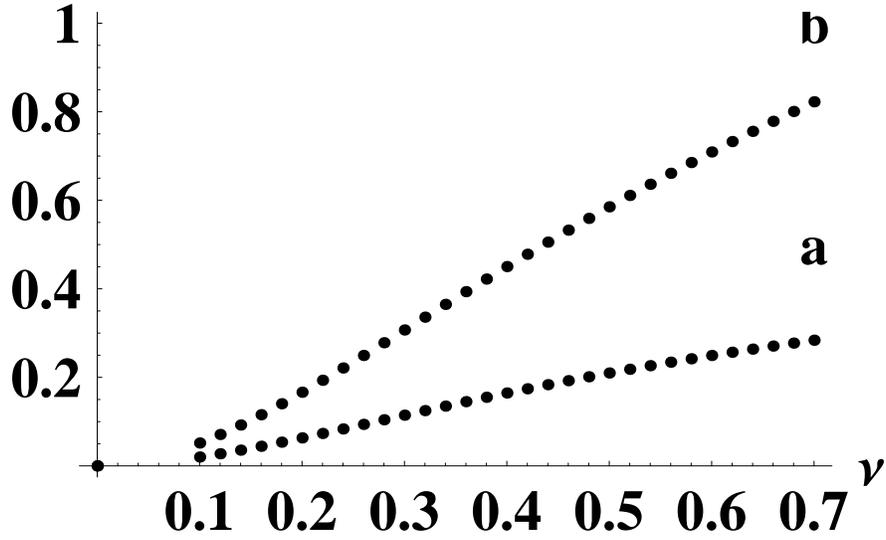}
\caption{\label{Fig:fig1} The $\nu$ - dependence of coefficients
 $a,b$ (see(19,20))}
\end{center}
\end{figure}

\begin{figure}
\begin{center}
\includegraphics[width=12cm]{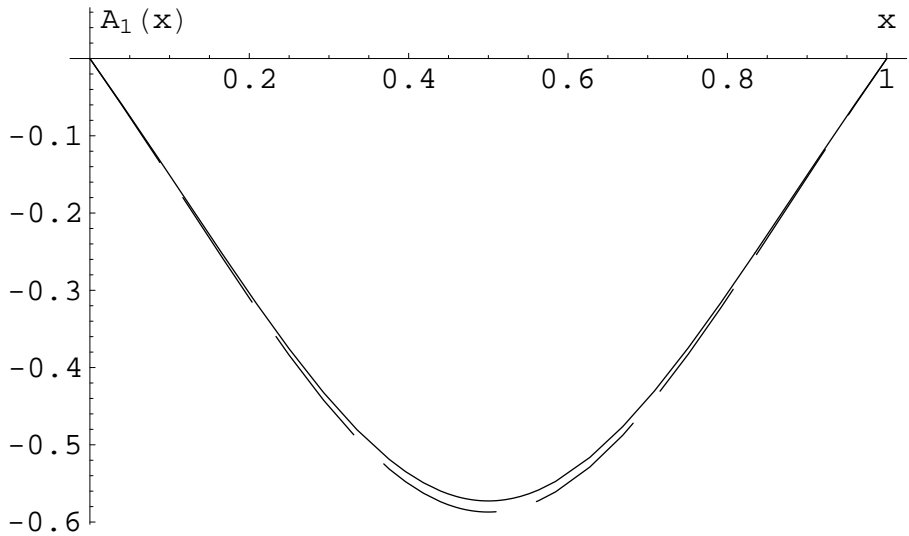}
\caption{Azimuthal asymmetry (see(26))}
\end{center}
\end{figure}

\section{Acknowledgements}
We are grateful to  V. Bytev, A. Korchin and E. Vinokurov for
discussions and help. Two of us (S. B., E. R.) acknowledge the
support of INTAS 05-1000 008-8328. Yu. P. and I. S. work was
supported by grant STCU-3239.

\section{Appendix}

For the case of screened potential the matrix element, corresponding
to two photon exchange have a form: \ba
M^{(2)}=\frac{s(4\pi\alpha)^{5/2}Z^2}{4\pi}N_P\sum^3_{i,j=1}\alpha_i\alpha_j
\bar{u}(p_2)\hat{R}\frac{\hat{P}}{s}v(p_1), \ea with \ba
N_P=\frac{1}{s}\bar{u}(P')\hat{k}u(P),\sum|N_P|^2=2, \ea and \ba
\hat{R}=\int\frac{d^2k_1}{\pi}\frac{1}{\vec{k}_1^2+\mu_i^2}\frac{1}{(\vec{q}-\vec{k}_1)^2+\mu_j^2}
\left[m\hat{e}S_2+2x\vec{T}_2\vec{e}+\hat{T}_2\hat{e}\right]; \nn \\
S_2=\frac{1}{c_1}-\frac{1}{c_{1k}}+\frac{1}{c_2}-\frac{1}{c_{2k}};
\quad
\vec{T}_2=\frac{\vec{p}_1}{c_1}+\frac{\vec{k}_1-\vec{p}_1}{c_{1k}}-
\frac{\vec{p}_2}{c_2}-\frac{\vec{k}_1-\vec{p}_2}{c_{2k}}, \ea where
$c_l$ was defined above, $c_{lk}=(\vec{k}_1-\vec{p}_l)^2+m^2$.

We need to calculate the integrals
\ba
B=\int\frac{d^2k_1}{\pi(\vec{k}_1^2+\mu^2_i)((\vec{q}-\vec{k}_1)^2+\mu^2_j)};\nn\\
(T_l,\vec{T}_l)=\int\frac{d^2k_1(1,\vec{k}_1)}{\pi(\vec{k}_1^2+\mu^2_i)((\vec{q}-\vec{k}_1)^2+\mu_j^2)
((\vec{p}_l-\vec{k}_1)^2+m^2)},
\ea
Applying Feynman joining procedure and performing standard Feynman parameter integration
we obtain:
\ba
T_l=\int^1_0dx\left\{-\frac{2A_1+B_1}{R\Delta}+\frac{2B_1}{R^{3/2}}
L \right\};\quad \Delta=A_1+B_1+C_1,\nn\\
\vec{T}_l=\int^1_0dx\vec{p}_{xl}\left\{\frac{2C_1+B_1}{R\Delta}-\frac{2C_1}{R^{3/2}}
L \right\}; \nn \\
L=ln\frac{(B_1+2C_1+\sqrt{R})^2}{4C_1\Delta},R=B_1^2-4A_1C_1>0,
\ea
with
\ba
A_1=-\vec{p}^2_{xl};\quad \vec{p}_{xl}=x\vec{q}+(1-x)\vec{p}_l; C_1=\mu_i^2;\nn\\
B_1=(1-x)(\vec{p}_l^2+m^2)+x(\vec{q}^2+\mu_j^2)-\mu_i^2.
\ea
In the kinematical region $m^2\sim \mu_i^2\ll\vec{q}^2\ll\vec{p}^2_i=\vec{p}_2^2=\vec{p}^2$
\ba
B=\frac{2}{\vec{q}^2}ln\frac{\vec{q}^2}{\mu_i\mu_j};\quad
T_1=T_2=\frac{2}{\vec{p}^2\vec{q}^2}ln\frac{\vec{q}^2}{\mu_i\mu_j};
\quad \vec{T}_1=\vec{T}_2=\frac{\vec{q}}{\vec{p}^2\vec{q}^2}ln\frac{\vec{p}^2}{\mu_j^2}.
\ea
In this limit we have $ S^{(2)}=\vec{T}^{(2)}=0$.
 Using the iteration procedure one can show as well
\be
  S^{(n)}=\vec{T}^{(n)}=0, n=2,3,4,...
\ee
providing the
absence of higher Coulomb corrections in this limiting case.

\end{document}